\documentstyle[12pt]{article}

\newcommand{\beq}{\begin{equation}}
\newcommand{\eeq}{\end{equation}}
\newcommand{\bea}{\begin{eqnarray}}
\newcommand{\eea}{\end{eqnarray}}

\newcommand{\gsim}{\lower.7ex\hbox{$
\;\stackrel{\textstyle>}{\sim}\;$}}
\newcommand{\lsim}{\lower.7ex\hbox{$
\;\stackrel{\textstyle<}{\sim}\;$}}

\def\op{{\bf P}}

\def\ot{{\bf T}}
\def\cp{{\bf CP}}
\def\cpt{{\bf CPT}}

\newcommand{\eod}{\end{document}}

\begin{document}
\thispagestyle{empty}
\vspace*{-20mm}

\begin{flushright}
UND-HEP-06-BIG\hspace*{.08em}06\\
hep-ph/0608073\\


\end{flushright}
\vspace*{7mm}

\begin{center}
{\LARGE{\bf
"I Know She Invented Fire, But  
\vspace*{2mm}
What Has She Done Recently?" -- 
\vspace*{2mm} 
On The Future Of Charm Physics}}
\footnote{Invited Lecture given at {\em CHARM 2006}, Beijing, June 2006}
\vspace*{14mm}

{\large{\bf I.I.~Bigi}}\\
\vspace{4mm}

 {\sl Department of Physics, University of Notre Dame du Lac}
\vspace*{-.8mm}\\
{\sl Notre Dame, IN 46556, USA}\vspace*{1.5mm}
email: ibigi@nd.edu

\vspace*{5mm}

{\bf Abstract}\vspace*{-.9mm}\\
\end{center}

\noindent
Detailed studies of weak charm decays fill an important {\em future} 
role in high energy physics. Chief among them are: (i) validating the  theoretical control achieved over hadronization as a worthwhile goal in its own right; (ii) calibrating our tools to saturate the discovery 
potential for New Physics in $B$ decays; (iii)  searching for New Physics in charm decays 
through hypothesis-generating research. The most promising area for the last item is a 
comprehensive study of \cp~violation. Since we need a new \cp~paradigm to  implement 
baryogenesis, this is not an idle goal. Charm decays provide opportunities unique among 
{\em up-}type quarks. While items (i) and (ii) will be addressed in a meaningful way and hopefully 
completed in the next few years, item (iii) will presumably require statistics that can be accumulated only by LHCb and a Super-B factory.

\vspace*{10mm}
\section{Prologue}

There is a common feeling charm physics had a great past -- it provided essential 
support for the paradigm shift to viewing quarks as physical degrees of freedom rather than 
objects of mathematical convenience -- yet it has no future. For the SM electroweak phenomenology 
of charm changing transitions appears on the decidedly dull side with the 
CKM parameters well-known due to three-family unitarity constraints, 
$D^0 - \bar D^0$ oscillations being slow, \cp~asymmetries small at best and loop driven decays extremely rare with huge backgrounds due to long distance dynamics. 

I do not view charm as a closed chapter.  Instead: 
"I have come to praise Ch., not to bury it!" 
To state it in more prosaic terms: there is a triple motivation for {\em further dedicated} studies 
of charm dynamics: 
\begin{itemize}
\item 
to gain new insights into nonperturbative dynamics and make progress in establishing theoretical 
control over them; 
\item 
to calibrate our theoretical tools for $B$ studies; 
\item 
to use charm transitions as a novel window onto New Physics.  
\end{itemize} 
Lessons from the first item will have an obvious impact on the tasks listed under the second and third  items. They might actually be of great value even beyond QCD, if the New Physics anticipated 
for the TeV scale is of the strongly interacting variety. 

\begin{figure}[t]
\vspace{5.0cm}
\includegraphics{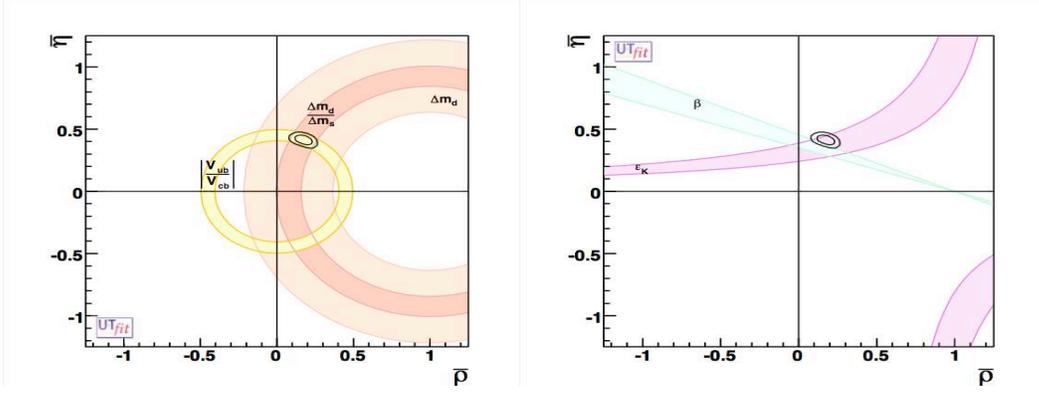}
 \caption{
      CKM unitarity triangle from $|V(ub)/V(cb)|$ and $\Delta M(B_d)/\Delta M(B_s)$ on the left and  
       compared to constraints from 
$\epsilon_K$ and sin2$\phi_1/\beta$ on the right 
(courtesy of M. Pierini)
    \label{fig1} }
\end{figure}

The accuracy of the theoretical description is of essential importance in this program. For we 
can{\em not 
count} on {\em numerically massive} interventions of New Physics in the decays of beauty 
mesons. This point is brought home again by the recently reported signal for $B_s - \bar B_s$ oscillations  \cite{D0BSOSC,CDFBSOSC}: 
\beq 
\Delta M(B_s) = 
\left\{   
\begin{array}{ll}
\left(19 \pm 2 \right) \, {\rm ps}^{-1} & {\rm D0} \\
\left(17.3 ^{+0.42}_{-0.21} \pm 0.07\right)\, {\rm ps}^{-1} & {\rm CDF} \\
\left(18.3 ^{+6.5}_{-1.5}\right) \, {\rm ps}^{-1} & {\rm CKM\; fit} \\
\end{array} 
\right. 
\label{DIRECTCPDATA} 
\eeq 
While the strength of the signal has not yet achieved 5 $\sigma$ significance, it looks most 
intriguing. {\em If} true, it represents another impressive triumph of CKM theory: 
the \cp~{\em in}sensitive observables $|V(ub)/V(cb)|$ and $\Delta M(B_d)/\Delta M(B_s)$ -- 
i.e. observables that do {\em not} require \cp~violation for acquiring a non-zero value -- imply 
\begin{itemize}
\item 
a non-flat CKM triangle and thus \cp~violation, see the left of Fig.~\ref{fig1} 
\item 
that is fully consistent with the observed \cp~sensitive observables $\epsilon_K$ 
and sin$2\phi_1$, see the right of Fig.~\ref{fig1}. 
\end{itemize}
My message is centered on three basic tenets: 

{\bf (i):} 
None of the new SM successes from the last few years weakens the case for 
New Physics even `nearby', namely around the TeV scale. 

{\bf (ii):} 
To learn about all the salient features of this anticipated New Physics we must study its impact on 
heavy flavour transitions -- even if it turns out in the end that none is observable. 
{\em \cp~studies are thus `instrumentalized' to probe for and analyze the New Physics, 
once it has emerged.}  

{\bf (iii):} 
We need 
precise, reliable and comprehensive studies of flavour dynamics; this means we have to look also at 
unusual places. 

For most details I refer the committed reader to several recent reviews 
\cite{BURD1,CICERONE,BURD2}. 

\section{The `Guaranteed' Profit }
\label{PROFIT}

\subsection{Lessons on QCD}
\label{LESSONS}

The issue at stake here is {\em not} whether QCD is the theory of the strong forces -- there is no alternative -- but our ability to perform calculations. Charm hadrons can act here as a bridge 
between the worlds of light flavours -- as carried by $u$, $d$ and $s$ quarks with masses lighter or at most comparable to $\Lambda _{QCD}$ and described by chiral perturbation theory -- and that of the 
bona fide heavy $b$ quark with $\Lambda_{QCD} \ll m_b$ treatable by heavy quark theory 
\cite{HQT}. 
\footnote{Top quarks have to be listed separately as `super-heavy', since due to 
$\Lambda_{QCD} \ll \Gamma_t$ they decay {\em before} they can hadronize \cite{RAPALLO}.}    

The verdict so far has been that charm acts `mostly somewhat' as a heavy quark: expansions in 
powers of $1/m_c$ basically work as far  as charm lifetimes are concerned, yet fail for light cone 
sum rules used to obtain the form factors for $D \to l \nu \pi/\rho$. The `a posteriori' explanation is that 
the latter contain corrections of order $1/m_c$ whereas the former start only at order $1/m_c^2$ 
\cite{CICERONE}. 

Quark models can serve as a most useful tool for training one's intuition and as a diagnostic of results 
from sum rules and lattice QCD; however, I do not view them as reliable enough for 
conclusive answers. 

Only lattice QCD carries the promise for a truly 
quantitative treatment of charm hadrons that can be improved {\em systematically}. 
Furthermore lattice QCD  is the only framework available that allows to approach charm from lower as well as higher mass scales, which involves different aspects of nonperturbative 
dynamics and thus -- if successful -- would provide impressive validation. 

Indulging myself in a short moment of bragging I would like to repeat what I had said in my talk 
at the 1993 Marbella Tau-Charm Workshop \cite{MARBELLA}: "The $\tau$-charm factory is the QCD machine 
for the 90's!" Ten years later the value of a such a factory was more widely appreciated, which 
led to the 
on-going CLEO-c and the future BESIII programs. At the same time we have to understand 
that the threshold for significance is much higher now than it was in the 1990's. 
This is due to a combination of several factors, chief among them the ability of the $B$ factories to perform high statistics as well as high  
quality charm studies  and the need for precision studies in $B$ decays. 

The required validation of lattice QCD has to go beyond a few `gold-plated' tests: even if it turns out 
that the measured value for the decay constant $f_D$ and the one inferred from lattice QCD were to 
agree within, say, a percent -- an impressive success for sure -- we can{\em not} conclude that 
there is a universal bound of a percent or two on the theoretical uncertainties even in 
semileptonic $D$ decays. Validation of lattice QCD requires accurate comparisons of the measured 
and predicted form factors in Cabibbo allowed as well as forbidden modes of 
$D^0$, $D^+$ and $D_s$ mesons. 

Charmonium studies provide yet another essential test ground; those are covered by other talks at this 
conference. 

\subsubsection{Is Charm Heavy?}

Let me list three pieces of evidence that charm is marginally heavy: 

{\bf (i)}: The value of the charm quark $\overline{MS}$ mass can be inferred from data also using methods other than 
lattice QCD \cite{BUCH}: 
\beq 
\overline m_c = 
\left\{   
\begin{array}{ll}
1.19 \pm 0.11 \; {\rm GeV} & {\rm charmonium \;  sum \; rules} \\
1.18 \pm 0.08 \; {\rm GeV} & {\rm moments \; of} \; B \to l \nu X 
\end{array} 
\right. 
\label{MC} 
\eeq 
The fact that two quite different theoretical treatments yield very consistent values supports  
that charm is somewhat heavy, i.e. significantly larger than $\Lambda_{QCD}$. 

{\bf (ii)}: More qualitative evidence is provided by the fact that the two channels 
$B \to l \nu D/D^*$ constitute about two thirds of the inclusive width for 
$B \to l \nu X_c$. For $D$ and $D^*$ form the ground states in heavy quark symmetry and have to 
saturate the inclusive semileptonic width for $m_c, m_b \to \infty$. 

{\bf (iii)}: As explained in detail in \cite{CICERONE} the lifetime ratios for the seven single charm hadrons that decay weakly are surprisingly well described by the heavy quark expansion; in some cases they were even predicted before data of the required accuracy existed. This fact appears quite 
nontrivial considering that these lifetime ratios span a factor of fourteen between the longest and the shortest  lifetimes, namely $\tau (D^+)$ and $\tau (\Omega_c)$. 

The SELEX collaboration has reported candidates for weakly decaying double-charm baryons. My 
judgment as a theorist is as follows: The reported lifetimes are way too short and do not exhibit the expected hierarchy \cite{CICERONE}. {\em If} SELEX's  interpretation is correct, then I had to conclude -- with obvious regret -- that the apparently successful description of single charm lifetimes was 
hardly more than a coincidence. 

\subsection{`Tooling up' for $B$ Studies}
\label{TOOLING}

Validating lattice QCD's result for $f_D$ and $f_{D_s}$ \cite{SHIPSEY06} would allow a rather trustworthy prediction for $f_B$ and $f_{B_s}$ by extrapolating $m_c \to m_b$. 
Yet there are many other applications for lessons learnt in charm decays. Some are 
obvious like extrapolating results on the form factors for $D_{(s)} \to l \nu \pi/K$ to 
$B_{(s)} \to l \nu \pi/K$, while others are not. I will give three examples of the latter. 

{\em Spectroscopy of open charm hadrons}: To extract $|V(cb)|$ and $|V(ub)|$ from inclusive 
semileptonic $B$ widths one needs to know the values of $m_b$, $m_c$ and other heavy quark parameters. Those are inferred from the shape of the lepton energy and hadronic mass  moments 
\cite{BUCH,GAMBURI}.  
In particular the latter are sensitive to the composition of the hadronic final state, the masses, widths 
and quantum numbers 
of the charm hadrons produced, i.e their spectroscopy. The limitations in our understanding of it 
\cite{ORSAY} at present represent one of the main systematic uncertainties. 
Assuming the wrong spectroscopy in the analysis could create a bias in the results inferred. 
The fact that a handful of heavy quark parameters describe so well a host of moments 
\cite{BUCH} with its many overconstraints shows that the SM $V-A$ currents dominate 
$B \to l \nu X$ \cite{SONG}; yet the aforementioned bias due to a wrong charm spectroscopy could  
hide the presence of non-SM chiralities or fake one in future studies. 

{\em Semileptonic $D$ decays}: In many models of New Physics there is a relatively clear 
connection between the \cp~violation observable in  
$B_d \to \phi K_S$ dominated by a strong Penguin operator and the rate for $B\to \gamma X_s$ 
given by the electroweak Penguin. Since the strength of the 
latter has been found to be close to the SM prediction, 
one finds rather tight bounds on the asymmetry in $B_d \to \phi K_S$. 
Yet there is an implicit 
assumption, namely that the emerging  photon is mostly {\em left}-handed as predicted by the SM. A 
non-SM contribution from a right-handed photon could not interfere with that from a left-handed one; 
thus it could contribute only quadratically to $\Gamma (B \to \gamma X_s)$. On the other hand 
the corresponding {\em strong} Penguin amplitude would interfere with the SM amplitude in the 
\cp~asymmetry  in $B_d \to \phi K_S$ thus contributing {\em linearly} and be of greater weight there. 

\noindent 
Measuring the photon polarization in radiative $B$ decays directly is a formidable task. It 
is more feasible to infer it indirectly from the exclusive mode $B \to \gamma K\pi \pi$ \cite{PIRJOL}. 
Most helpful or even essential information on the dynamical structure of the relevant 
$K\pi \pi$ system can be obtained by analyzing the semileptonic charm channel 
$D \to l \nu K\pi \pi$.  

{\em (Time dependent)  Dalitz plot studies}: The most intriguing indication for New Physics in 
heavy flavour decays has emerged in the time-dependent \cp~asymmetries for 
$B_d \to \phi K_S$ (and related channels). A reliable SM prediction tells us that it should 
closely mirror the situation of $B_d \to \psi K_S$ with the same coefficients for the 
sin \& cos$\Delta M_dt$ terms: $S \simeq 0.68$ \& $C\simeq 0$. The values of the 
$S$ term measured by BELLE and BABAR \cite{BSSS} -- while not inconsistent with the SM expectations -- 
are on the low side by an amount that would be natural for New Physics. Future analyzes might 
turn this into a significant discrepancy. 

\noindent Yet one has to be aware of the following complication: One has to extract 
$B_d \to \phi K_S$ from $B_d \to K^+K^- K_S$. While the $\phi$ (in contrast to the $\rho$) 
represents a rather narrow resonance, it still has a finite width. Making merely a mass cut 
on the kaon pair will let other contributions `slip through', non-resonant ones or other  
resonances like the scalar $f(980)$. If $K^+K^-$ form a {\em scalar} pair, then the final state 
in $B_d \to [K^+K^-]_{J=0}K_S$ has the opposite 
\cp~parity than $B_d \to \phi K_S$, and it will have a \cp~asymmetry equal in magnitude, yet 
{\em opposite in sign} to that of $B_d \to \phi K_S$ (if driven by the same quark level 
operator). To give an example for {\em illustration}: 
a $B_d \to [K^+K^-]_{J=0}K_S$ {\em amplitude} 10\% the size 
of the dominant $B_d \to \phi K_S$ amplitude in the sample would {\em reduce} the 
\cp~asymmetry by 20\% -- i.e. significantly. A detailed analysis of 
{\em time-dependent Dalitz plots} will allow to disentangle such effects. 
This comes with a hefty price of course, namely that of requiring huge statistics. 
Yet, adapting a quote from Greek antiquity:  "There is no royal way to fundamental insights."  

\noindent 
Finally and most importantly for this talk, one can learn many lessons about hadronization and final state interactions relevant for $B$ decays by studying the corresponding Dalitz plots for charm decays like $D \to 3 K$. For most of the clear resonance structures lies below 1.5 GeV; also to first approximation the excitation curve for a resonance $R$ produced in $D$ or $B \to RM$ with $M$ 
denoting a light flavour meson should be very similar. I would like to add, however, that BELLE 
data on $B \to K \pi \pi$ indicate that these are not absolute rules \cite{BELLE0412}.

\section{`The Best might still be ahead'}
\label{BEST} 

There are two kinds of research, namely `hypothesis-driven' and `hypothesis-generating' 
research. The first kind is essential -- and favoured by funding agencies. Yet also the 
second kind -- `thinking outside the box' -- must be pursued, although it is much harder to plan; we owe many of the fundamental paradigm shifts to such an approach.  The program of the $B$ factories has 
been largely of the `hypothesis-driven' variety, and a most successful one at that. 

The situation is quite different with charm dynamics. Charm spectroscopy has led to the recent 
renaissance in `hypothesis-generating'  studies of QCD. The best {\em long}-term motivation for 
a future charm program is a `hypothesis-generating' search for New Physics. 
To use an analogy from real life: "If baseball teams from Boston 
and Chicago can win the World Series in two successive years -- overcoming curses having lasted more than 80 years -- then charm can surely reveal New Physics." 

New Physics scenarios in general 
induce flavour changing neutral currents that {\em a priori} have little reason to be as much suppressed as in the SM. More specifically they could be substantially stronger for up-type than for down-type quarks; 
this can happen in particular in models which have to reduce strangeness changing neutral currents 
below phenomenologically acceptable levels by some alignment mechanism. 

In such scenarios charm plays a unique role among the up-type quarks $u$, $c$ and $t$; for only 
charm allows the full range of probes for New Physics in general and flavour-changing 
neutral currents in particular: 
(i) Since top quarks do not hadronize \cite{RAPALLO}, there can be no $T^0- \bar T^0$ oscillations. More generally, hadronization, while hard to bring under theoretical control, enhances the 
observability of \cp~violation. 
(ii)  
As far as $u$ quarks are concerned, $\pi^0$, $\eta$ and $\eta ^{\prime}$ decays electromagnetically, not weakly. They are their own antiparticles and thus cannot oscillate. \cp~asymmetries are mostly 
ruled out by \cpt~invariance. 

My basic contention can then be formulated as follows: {\em Charm transitions provide a unique portal 
for a novel access to flavour dynamics with the experimental situation being a priori quite 
favourable (apart from the absence of Cabibbo suppression). Yet even that handicap can be overcome 
by statistics. }

\subsection{`Inconclusive' $D^0 - \bar D^0$ Oscillations}
\label{DOSC}

$D^0 - \bar D^0$ oscillations can be characterized as follows: 

\noindent $\oplus$ They represent a fascinating quantum mechanical problem; 

\noindent $\ominus$ while they provide only an ambiguous probe for New Physics, 

\noindent $\oplus$ they are an important ingredient in \cp~asymmetries that, if observed, would establish 
the intervention of New Physics. 

Oscillations are characterized by two dimensionless ratios: 
\beq 
x_D \equiv \frac{\Delta M_D}{\Gamma _D} \; , \; y_D \equiv \frac{\Delta \Gamma_D}{2\Gamma _D} 
\eeq
A conservative rather model independent bound reads $x_D$, $y_D$ $\leq {\cal O}(0.01)$ \cite{CICERONE}.  With 
present data reading \cite{ASNER} 
\beq 
\left. x_D\right| _{exp} <  0.03 \; , \; \left.  y_D\right| _{exp} \sim 0.01 \pm 0.005
\eeq
one can conclude that a meaningful search for $D^0$ oscillations has `only just' begun. 

At this point allow me a personal comment: the (in)famous `Nelson plot' 
\cite{NELSON} on theoretical predictions 
concerning $x_D$ was witty and an appropriate reminder for theorists to use some common 
sense. Yet now it should be retired with honour, since we have a considerably better understanding 
of the dynamical issues involved.  

It is widely understood that the usual quark box diagram is utterly irrelevant due to its untypically severe 
GIM suppression $(m_s/m_c)^4$. 
A systematic 
analysis based on an OPE has been given \cite{BUDOSC} in terms of powers of 
$1/m_c$ and $m_s$. Contributions from higher-dimensional operators with a much softer 
GIM reduction of $(m_s/\mu_{had})^2$ due to `condensate'  terms in the OPE  yield 
\beq 
\left. x_D (SM)\right|_{OPE}, \; \left. y_D (SM)\right|_{OPE} \sim {\cal O}(10^{-3}) \; . 
\label{XDYDPRED}
\eeq 
The authors of \cite{FALK} find very similar numbers, albeit in a quite different approach. 
When evaluating the predictions in Eq.\ref{XDYDPRED} one has to distinguish carefully 
between two similar sounding questions: 
\begin{itemize}
\item 
"What are the {\em most likely} values for $x_D$ and $y_D$ within the SM?" 

My answer as given above: For both $\sim {\cal O}(10^{-3})$. 
\item 
"How large could $x_D$ and $y_D$ {\em conceivably} be within the SM?" 

My answer: One cannot rule out $10^{-2}$. 
\end{itemize}
While one predicts similar numbers for $x_D(SM)$ and $y_D(SM)$, one should note 
that they arise in very different dynamical environments. $\Delta M_D$ being generated from 
{\em off}-shell intermediate states is sensitive to New Physics, which could produce 
$x_D \sim {\cal O}(10^{-2})$. $\Delta \Gamma_D$ on the other hand is shaped by 
{\em on}-shell intermediate 
states; while it is hardly sensitive to New Physics, it involves much less averaging or `smearing' than 
$\Delta M_D$ making it thus more vulnerable to violations of quark-hadron duality. 
\footnote{A similar concern applies to $\Delta \Gamma (B_s)$.}  
Observing 
$y_D \sim 10^{-3}$ together with $x_D \sim 0.01$ would provide intriguing, though not conclusive 
evidence for New Physics, while $y_D \sim 0.01 \sim x_D$ would pose a true conundrum for its 
interpretation. 

This skepticism does not mean one should not make the utmost efforts to probe 
$D^0 - \bar D^0$ oscillations down to the $x_D$, $y_D$ $\sim 10^{-3}$ level. For one we might be only one theory breakthrough away from making a 
precise prediction. Yet more importantly this challenge provides an important 
experimental validation check when searching for a \cp~asymmetry involving oscillations.

\subsection{\cp~Violation with \& without Oscillations}
\label{CPV}

Most factors favour dedicated searches for \cp~violation in charm transitions: 

 $\oplus$ 
Since baryogenesis implies the existence of New Physics in \cp~violating dynamics, it would be unwise not to undertake dedicated searches for \cp~asymmetries in 
charm decays, where the `background' from known physics is between absent and small: 
for within the SM the effective weak phase is highly diluted, namely $\sim {\cal O}(\lambda ^4)$, and it can arise only in {\em singly Cabibbo suppressed} transitions, where one  
expects asymmetries to reach the ${\cal O}(0.1 \%)$ level; significantly larger values would signal New Physics.  
{\em Any} asymmetry in {\em Cabibbo 
allowed or doubly suppressed} channels requires the intervention of New Physics -- except for 
$D^{\pm}\to K_S\pi ^{\pm}$ \cite{CICERONE}, where the \cp~impurity in $K_S$ induces an asymmetry of $3.3\cdot 10^{-3}$. One should keep in mind that in going from Cabibbo allowed to Cabibbo 
singly and doubly  suppressed channels, the SM rate is {\em suppressed} by factors of about 
twenty and four hundred, respectively: 
$$ 
\Gamma _{SM}( H_c \to [S=-1]) : \Gamma _{SM}( H_c \to [S= 0]) : \Gamma _{SM}( H_c \to [S= +1]) 
\simeq 
$$
\beq
1 : 1/20 : 1/400
\eeq
One would expect that this suppression will enhance the visibility of New Physics.  

$\oplus$ 
Strong phase shifts 
required for {\em direct} \cp~violation to emerge in partial widths are in general large as are the branching ratios into relevant modes;  while large final state interactions complicate the 
interpretation of an observed signal in terms of the microscopic parameters of the underlying dynamics, they enhance its observability.  

$\oplus$ 
\cp~asymmetries can be linear in New Physics amplitudes thus increasing sensitivity to the 
latter.  

 $\oplus$ 
Decays to final states of {\em more than} two pseudoscalar or one pseudoscalar and one vector meson contain 
more dynamical information than given by their  widths; their distributions as described by Dalitz plots 
or \ot~odd moments can exhibit \cp~asymmetries that might be considerably larger than those for the 
width. Final state interactions while not necessary for the emergence of such effects, can fake a signal; 
yet that can be disentangled by comparing \ot~odd moments for \cp~conjugate modes \cite{PEDRINI}: 
\beq 
O_T(D\to f) \neq - O_T(\bar D \to \bar f) \; \; \; \Longrightarrow \; \; \; \cp~{\rm violation}
\eeq
I view this as a very promising avenue, where we still have to develop the most effective analysis tools for small 
asymmetries. Below I will briefly illustrate the general method by one explicit example. 

 $\oplus$ The distinctive channel $D^{\pm*} \to D \pi^{\pm}$ provides a powerful tag 
on the flavour identity of the neutral $D$ meson. 

 $\ominus$ The `fly in the ointment' is that $D^0 - \bar D^0$ oscillations are on the slow side.

$\oplus$ Nevertheless one should take on this challenge. For 
\cp~violation involving $D^0 - \bar D^0$ oscillations is a reliable probe of New Physics: the 
asymmetry is controlled by  
sin$\Delta m_Dt$ $\cdot$ Im$(q/p)\bar \rho (D\to f)$. Within the SM both factors are small, namely 
$\sim {\cal O}(10^{-3})$, making such an asymmetry unobservably tiny -- unless there is 
New Physics; for a recent New Physics model see \cite{PEREZ}.  
One should note 
that this observable is {\em linear} in $x_D$ rather than quadratic as for \cp~insensitive quantities 
like $D^0(t) \to l^-X$.  
$D^0 - \bar D^0$ oscillations, \cp~violation and New Physics might thus be discovered simultaneously in a transition. Such effects can be searched for in final states common to $D^0$ 
and $\bar D^0$ decays like \cp~eigenstates -- $D^0 \to K_S\phi$, $K^+K^-$, $\pi^+\pi^-$ -- or 
doubly Cabibbo suppressed modes -- $D^0 \to K^+\pi^-$. In the end it might turn out that the 
corresponding three-body final states -- $D^0 \to K_S \pi^+\pi^-$, $D^0 \to K^+K^-\pi^0/\pi^+\pi^-\pi^0$ 
and $D^0 \to K^+\pi^- \pi^0$ -- allow searches with higher sensitivity. Undertaking 
{\em time-dependent} Dalitz plot studies requires a higher initial overhead, yet in the long run this 
should pay handsome dividends exactly since Dalitz analyses can invoke many internal correlations 
that in turn serve to control systematic uncertainties. 

$\oplus$ It is all too often overlooked that \cpt~invariance can provide nontrivial constraints on 
\cp~asymmetries. For it imposes equality not only on the masses and total widths of particles and antiparticles, but also on the widths for `disjoint' {\em sub}sets of channels. 
`Disjoint' subsets are the decays to final states that can{\em not} rescatter into each other. Examples are 
semileptonic vs. nonleptonic modes with the latter subdivided further into those with strangeness 
$S = -1,0.+1$. Observing a \cp~asymmetry in one channel one can then infer in which other channels 
the `compensating' asymmetries have to arise \cite{CICERONE}. 

\subsubsection{Theoretical Engineering}
\label{ENGIN}

\cp~asymmetries in integrated partial widths depend on hadronic matrix elements and (strong)  
phase shifts, neither of which can be predicted accurately. However the craft of theoretical 
engineering can be practiced with profit here. One makes an ansatz for the general form of the matrix 
elements and phase shifts that are included in the description of $D\to PP, PV, VV$ etc. 
channels, where $P$ and $V$ denote pseudoscalar and vector mesons, and fits them to the measured branching ratios on the Cabibbo allowed, once and twice forbidden level. If one has sufficiently accurate and comprehensive data, one can use these fitted values of the hadronic parameters to predict \cp~asymmetries. Such analyses have been undertaken in the past \cite{LUSIG}, 
but the data base was not as broad and precise as one would like. 
{\em CLEOc and BESIII measurements 
will certainly lift such studies to a new level of reliability.}   

\subsubsection{An Example for a \ot~odd Correlation}
\label{DKKPIPI}

\cp~asymmetries in final state distributions can be substantially larger than in integrated partial 
widths. A dramatic example for that has been found in $K_L$ decays. Consider the 
rare mode $K_L \to \pi^+\pi^- e^+e^-$ and define by $\phi$ the angle between the 
$\pi^+\pi^-$ and $e^+e^-$ planes. The differential width has the general form 
\beq 
\frac{d\Gamma}{d\phi}(K_L \to \pi^+\pi^- e^+e^-) = \Gamma_1 {\rm cos}^2 \phi + 
\Gamma_2 {\rm sin}^2 \phi + \Gamma_3 {\rm cos} \phi {\rm sin}\phi
\eeq
Upon integrating over $\phi$ the $\Gamma_3$ term drops out from the total width, which thus is 
given in terms of $\Gamma_{1,2}$ with $\Gamma_3$ representing a forward-backward asymmetry. 
\beq 
 \langle A \rangle \equiv 
 \frac{\int _0 ^{\pi/2}\frac{d\Gamma}{d\phi} -  \int _{\pi/2} ^{\pi}\frac{d\Gamma}{d\phi}}
{\int _0 ^{\pi}\frac{d\Gamma}{d\phi} } = 
\frac{2\Gamma_3}{\pi(\Gamma_1 + \Gamma_2)}
\eeq
Under \op~and \ot~one has cos$\phi$sin$\phi$ $\to$ - cos$\phi$ sin$\phi$. 
Accordingly $A$ and $\Gamma_3$ constitute a \ot~odd correlation, while $\Gamma_{1,2}$ are 
\ot~even. $\Gamma_3$ is driven by the \cp~impurity $\epsilon_K$ in the kaon wave function. 
$\langle A \rangle$ has been measured to be large in full agreement with theoretical 
predictions \cite{SEHGALWANN}: 
\beq
 \langle A \rangle = 0.138 \pm 0.022  \; . 
\eeq
One should note this observable is driven by $|\epsilon_K| \simeq 0.0023$. 

$D$ decays can be treated in an analogous way.  Consider the Cabibbo suppressed channel 
\footnote{This mode can exhibit direct \cp~violation even within the SM.}
\beq 
\stackrel{(-)}D \to K \bar K \pi^+\pi^-
\eeq
and define by $\phi$ now the angle between the $K \bar K$ and $\pi^+\pi^-$ planes. Then 
one has 
\bea 
\frac{d\Gamma}{d\phi}(D \to K \bar K\pi^+\pi^-) &=& \Gamma_1 {\rm cos}^2 \phi + 
\Gamma_2 {\rm sin}^2 \phi + \Gamma_3 {\rm cos} \phi {\rm sin}\phi \\
\frac{d\Gamma}{d\phi}(\bar D \to K \bar K\pi^+\pi^-) &=& \bar \Gamma_1 {\rm cos}^2 \phi + 
\bar \Gamma_2 {\rm sin}^2 \phi + \bar \Gamma_3 {\rm cos} \phi {\rm sin}\phi 
\eea
As before the partial width for $D[\bar D] \to K\bar K \pi^+\pi^-$ is given by 
$\Gamma_{1,2} [\bar \Gamma_{1,2}]$; $\Gamma_1 \neq \bar \Gamma_1$ or 
$\Gamma_2 \neq \bar \Gamma_2$ represents direct \cp~violation in the partial width. 
$\Gamma_3 \& \bar \Gamma_3$ constitute \ot~odd correlations. By themselves they do not necessarily 
indicate \cp~violation, since they can be induced by strong final state interactions. However 
\beq 
\Gamma_3 \neq \bar \Gamma_3 \; \; \Longrightarrow \cp~{\rm violation!}
\eeq 
It is quite possible or even likely that a difference in $\Gamma_3$ vs. $\bar \Gamma_3$ 
is significantly larger than in $\Gamma_1$ vs. $\bar \Gamma_1$ or 
$\Gamma_2$ vs. $\bar \Gamma_2$. Furthermore one can expect that differences in detection 
efficiencies can be handled by comparing $\Gamma_3$ with $\Gamma_{1,2}$ and 
$\bar \Gamma_3$ with $\bar \Gamma_{1,2}$.



\subsubsection{Experimental Status \& Future Benchmarks}
\label{BENCH}

Time integrated \cp~asymmetries have been analyzed where sensitivities of order 1\% 
[several \%] have been achieved for Cabibbo allowed and once suppressed modes with two 
[three] body final states \cite{SHIPSEY06}. Time {\em dependent} \cp~asymmetries (i.e. those involving 
$D^0 - \bar D^0$ oscillations) still form largely `terra incognita'. 

Since the primary goal is to establish the intervention of New Physics,   
one `merely' needs a sensitivity level above the reach of the SM; `merely' does not mean 
it can easily be achieved. As far as {\em direct} \cp~violation is concerned -- 
in partial widths as well as in final state distributions -- this means asymmetries down to the 
$10^{-3}$ or even $10^{-4}$ level in  Cabibbo allowed channels and 1\% level or better 
in twice Cabibbo suppressed modes;  in Cabibbo once suppressed decays one wants 
to reach the $10^{-3}$ range although CKM dynamics can produce effects of that order 
because future advances might sharpen the SM predictions -- and one will get them 
along with the other channels. For  
{\em time dependent} asymmetries in $D^0 \to K_S\pi^+\pi^-$, $K^+K^-$, $\pi^+\pi^-$ etc. 
and in $D^0 \to K^+\pi^-$ 
one should strive for the  
${\cal O}(10^{-4})$ and ${\cal O}(10^{-3})$ levels, respectively. 

Statisticswise these are not utopian goals considering that LHCb expects to record about 
$5 \cdot 10^7$ {\em tagged} $D^* \to D + \pi  \to K^+K^- +\pi$ events in a nominal year 
of $10^7$ s \cite{TAT}. 

When going after asymmetries below the 1\% or so level one has to struggle against systematic uncertainties, in particular since detectors are made from matter. I can see three powerful weapons in this struggle: 
(i) 
Resolving the time evolution of asymmetries that are controlled by $x_D$ and $y_D$, which requires 
excellent microvertex detectors; (ii)  Dalitz plot consistency checks;  
(iii) quantum statistics constraints on distributions, \ot~odd moments etc. \cite{QCORREL}

\section{Outlook -- not an Epilogue}
\label{OUT}

We still have two truly central tasks to address in charm studies. 

$\bullet$  
To validate the {\em quantitative theoretical} control one has or soon will achieve over 
hadronization:   (a) it is valuable in its own right (and extending such studies to charm baryons would provide us with novel perspectives onto nonperturbative QCD), and (b) it    
will sharpen our tools for $B$ decay studies to saturate the discovery potential for New Physics 
there.  

$\bullet$ 
{\em Unique} searches for New Physics with {\em up}-type quarks:  
(a) While probing $D^0 - \bar D^0$ oscillations represents an important intermediate stage, 
searches for \cp~violation are the essential goal. We should not forget that a new 
\cp~paradigm is needed for baryogenesis. 
(b) 
The experimental situation is mostly favourable in the charm sector,  
yet even so we need as much statistics as possible; it will be most desirable that 
LHCb can contribute to detailed charm studies and that a Super-$B$ factory will be 
realized.  
(c) BESIII will make important contributions, yet not provide final answers. 

\noindent 
While no evidence for New Physics has so far been found in charm decays, we should not 
get discouraged: for only recently have we entered a domain where one could 
{\em realistically hope} for an effect.

\section*{Acknowledgments} It is a pleasure to thank the organizers for creating a fine 
workshop and being such gracious hosts. I also enjoyed and benefitted from many stimulating 
discussions during this workshop, in particular with D. Asner, A. Bondar, HaiBo Li, 
A. Palano and I. Shipsey. This work was supported by the NSF under grant PHY03-55098.


\end{document}